# Common Coil Dipole for High Field Magnet Design and R&D

March 16, 2022

Ramesh Gupta, Kathleen Amm, Julien Avronsart, Michael Anerella, Anis Ben Yahia, John Cozzolino, Piyush Joshi, Mithlesh Kumar, Febin Kurian, Chris Runyan, William Sampson, Jesse Schmalzle, Brookhaven National Laboratory, Upton, NY 11973, USA

Stephan Kahn, Ronald Scanlan, Robert Weggel, Erich Willen, Particle Beam Lasers, Inc., 8800 Melissa Court, Waxahachie, TX 75167-7279, USA

Qingjin Xu, Institute of High Energy Physics, Institute of High Energy Physics, Chinese Academy of Sciences, Beijing, China

Javier Munilla, Fernando Toral, Centro de Investigaciones Energéticas, Medioambientalesy Tecnológicas (CIEMAT), Avda. Complutense, 22, E-28040 Madrid, Spain

Paolo Ferracin, Steve Gourlay, GianLuca Sabbi, Xiaorong Wang, Lawrence Berkeley National Laboratory, 1 Cyclotron Road, Berkeley, CA 94720, USA

Danko van der Laan, Jeremy Weiss, Advanced Conductor Technologies LLC, 2200 Central Avenue, Suite A/B, Boulder, CO 80301, USA

Email: gupta@bnl.gov

## Introduction

The principle thrust of magnet R&D for the next generation high energy hadron collider is developing and demonstrating magnet designs for building higher field magnets in large numbers in industry at a lower cost. All accelerators to date have been built on the strength of the Cosine Theta (CT) magnet designs using NbTi and, more recently, $Nb_3Sn$ Low Temperature Superconductors (LTS). These magnet designs are being extended to reach higher fields. Based on a significant body of experience, the cost of carrying out large-scale production based on this design as well as the challenges in obtaining higher fields both in terms of budget and turn-around-time can be estimated. With the next collider several decades away, a case can be made that this time should be used in demonstrating alternate designs that have the potential to produce higher field magnets in industry at a lower cost while meeting the field quality requirements of high energy colliders. There is also a need to include a strong R&D component in the program that facilitates testing and demonstration of new conductors and new technologies in a relatively shorter time frame and at a cost much smaller than building a complete prototype magnet.

The common coil design [1] is a conductor-friendly block coil design with simple ends that have a large bend radius (see Fig. 1). The bend radius is determined by the separation between the two apertures of the collider rather than the aperture itself. The common coil design easily accommodates high field, brittle conductors or those cables that require large bend radii. The large bend radii in the common coil geometry allows both "Wind & React" and "React & Wind" technologies. In several ways, the common coil design may be a technically superior solution for high field magnets because the coils are primarily stacked vertically and move as a unit against the large horizontal Lorentz forces. This largely eliminates the internal strain on the conductor at or near the end region of the superconducting coils when the two sides of the coil move apart under Lorentz forces – a very different situation as compared to that in the conventional block coil or CT dipole designs. As compared to the conventional block coil designs, the common coil block coil design eliminates almost all of the hard-way bends (or ends requiring long length). The only remaining hard-way bends are in the small pole coils (see Fig. 1, right), and some designs practically eliminate the hard-way bends in those as well.

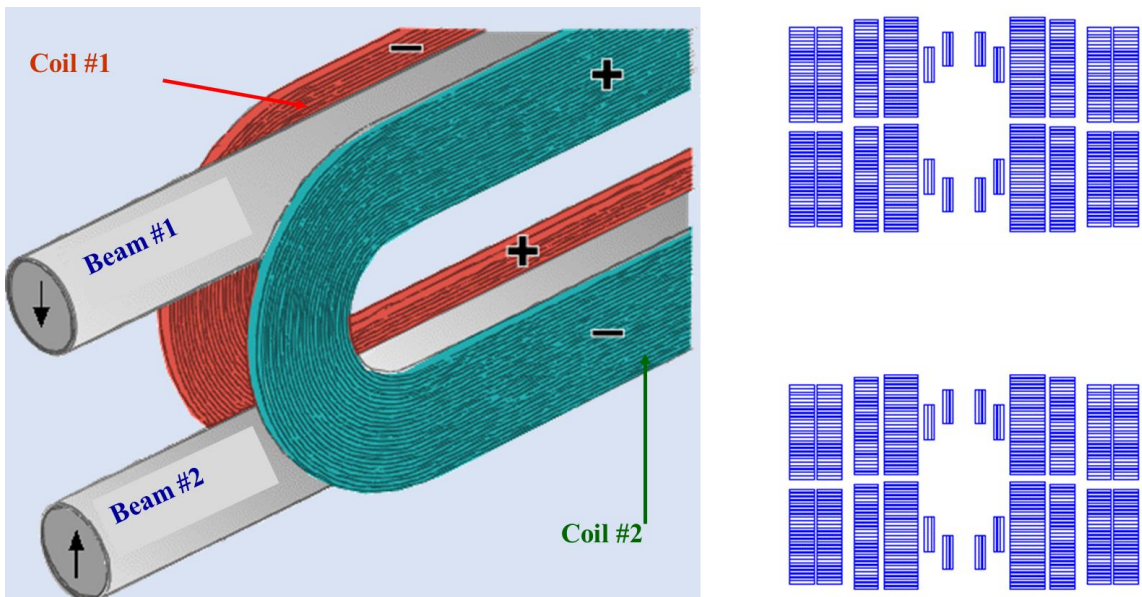

*Fig. 1: Main coils of the common coil design (left), main coils (horizontally oriented) with the pole coils (vertically oriented) providing the field quality.*

The common coil design facilitates a modular geometry that is particularly attractive for hybrid (HTS/LTS) magnet designs that require combining coils made with different types of conductors. In addition, the common coil design also offers easier vertical segmentation which is ideally suitable for hybrid coil dipole designs. Such magnets use coil modules made with different conductors ($Nb_3Sn$, NbTi and HTS). The segmentation between HTS and LTS is efficient in common coil dipole because it has been found that only one HTS coil (in addition to the pole coil) would be sufficient for the highest dipole field under consideration (20 T with 15% operational margin). Moreover, the design also provides natural and easier stress management. These features are applicable for both R&D magnets and for large-scale production magnets.

In addition to allowing versatility in conductors and technologies, the common coil design is also one of the most likely candidates to provide lower cost large scale production of high field 2-in-1 collider dipoles with good technical performance. Lower cost in large volume industrial manufacturing is expected because the common coil design would be less expensive and could employ more reliable production techniques due to 1) its simple racetrack coil geometry; 2) half the number of coils required (as the same coils are shared between two apertures), and 3) the geometry requires less structural material.

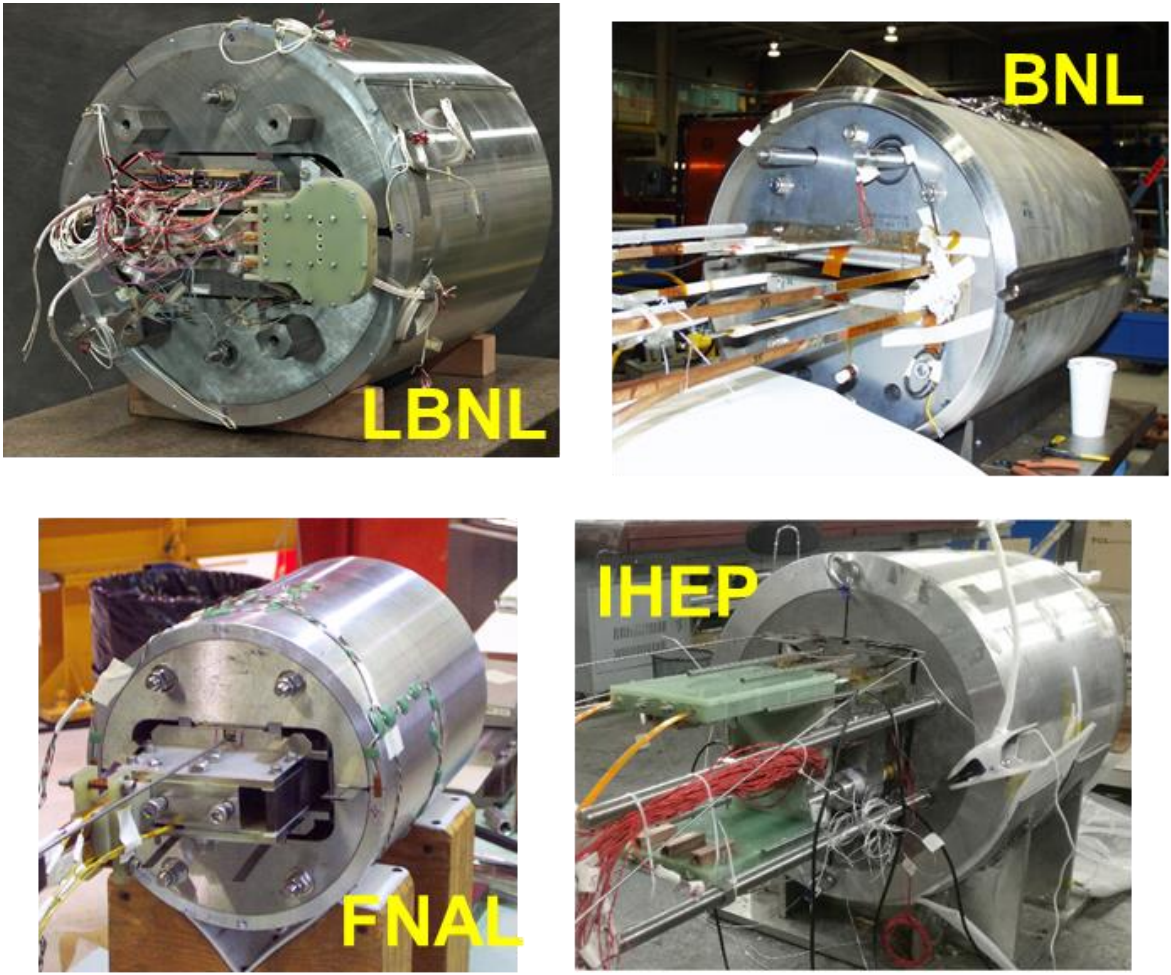

*Fig. 2: R&D common coil dipoles built and tested at LBNL, BNL, FNAL and IHEP.*

## Previous work on the common coil design

The common coil dipole design was used in an earlier proposal for the Very Large Hadron Collider (VLHC) in the United States (US) [2]. The common coil design has also been used in the present proposal of the Super proton-proton Collider (SppC) in China [3] and is one of the designs under consideration for the proposed Future Circular Collider (FCC) by CIMET [4].

Several institutes including LBNL [5], BNL [6], FNAL [7], IHEP [3] and CERN collaborators at CIEMAT [4] have carried out significant design studies on common coil magnets. Magnets based on the common coil dipole design have been successfully built (see Fig. 2) at Lawrence Berkeley National Laboratory (LBNL), Brookhaven National Laboratory (BNL), Fermilab National Accelerator Laboratory (FNAL), and Institute of High Energy Physics (IHEP) and other institutions with a variety of superconductors such as NbTi, $Nb_3Sn$, Bi-2212, ReBCO and Iron Based Superconductors (IBS). The very first test magnet based on this design at LBNL (RT1) reached short sample with almost no quenches [8]. Similar results were obtained at many other institutions including at BNL [9]. Further tests also showed that the change in pre-stress causes no degradation in performance. A common coil $Nb_3Sn$ dipole (RD3) built using the "Wind & React" technology reached 14.7 T at LBNL [5]. At BNL a "React & Wind" $Nb_3Sn$ dipole DCC017 [6] was built with essentially no vertical and horizontal pre-stress and it reached over 10 T, its computed short sample limit. FNAL [7] and IHEP [10] have also built and tested magnets based on this design. Despite many successes and even though FNAL built and tested an accelerator type field quality common coil dipole, demonstration of a fully optimized, high quality, high field common coil design with reasonable aperture and good technical performance remains to be done.

Hybrid magnets based on the common coil design have also been built and tested. Some examples of a hybrid common coil dipole include a 3.7 T dipole with Bi-2212 and $Nb_3Sn$ Rutherford cables [11] at BNL, a 10.7 T dipole at 4.2 K with $Nb_3Sn$ and NbTi cables at IHEP [10], an 8.7 T dipole with ReBCO tape in a perpendicular direction to the primary field and $Nb_3Sn$ cable at BNL in collaboration with Particle Beam Lasers, Inc. [12], and a 12.3 T dipole with ReBCO tape in primary field parallel direction and $Nb_3Sn$ cable at BNL [13]. Higher field hybrid common coil dipoles under construction include 13-14 T with CORC® cable and $Nb_3Sn$ cable under a collaboration between the Advanced Conductor Technologies, LLC and BNL [14], and HTS (ReBCO or IBS) at IHEP [15]. A common feature of all these dipoles is easy and better optimized segmentation between coils made with more than one type of conductor.

## 16 T $Nb_3Sn$ common coil design study for FCC

A common coil design option was examined as a part of the design study for 16 T $Nb_3Sn$ common coil dipole for FCC by CIMET [4]. Fig. 3 shows the magnetic and mechanical design.

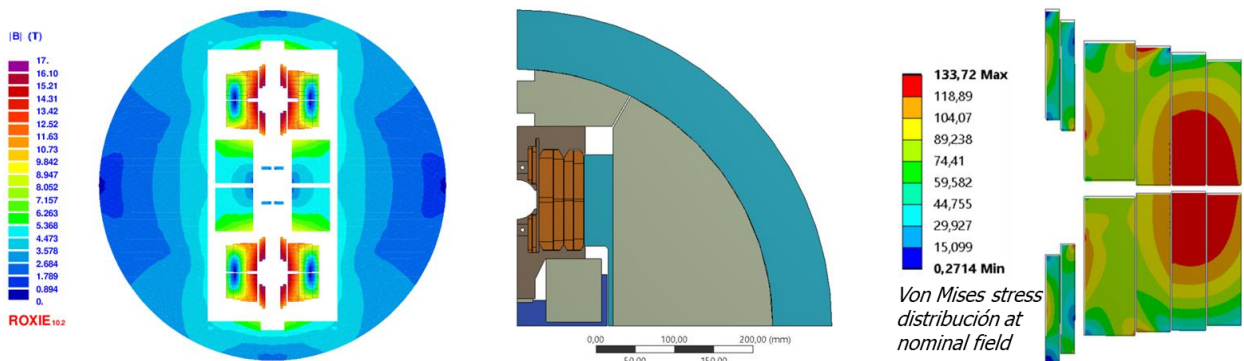

*Fig. 3: Design studies performed for 16 T common coil dipole for FCC by CIMET.*

The design of a 16 T Nb$_3$Sn common coil dipole was also carried out by Particle Beam Lasers, Inc. (PBL) and the BNL team as a part of a Small Business Innovation Research (SBIR) program. The optimized coil design is shown in Fig. 4 (left). It has less than 0.3% peak enhancement (maximum field on the conductor with respect to the field at the center of the bore). Computed harmonics at the design field of 16 T remain less than three units at the design field of 16 T. Fig. 4 (center) shows one candidate layout of the coils and Fig. 4 (right) shows an ANSYS model showing the structure considered and Lorentz forces applied. Computed Von Mises stresses and strains in the main coils and pole coils are shown in Fig. 5. They remain within the acceptable limit for the Nb$_3$Sn coils.

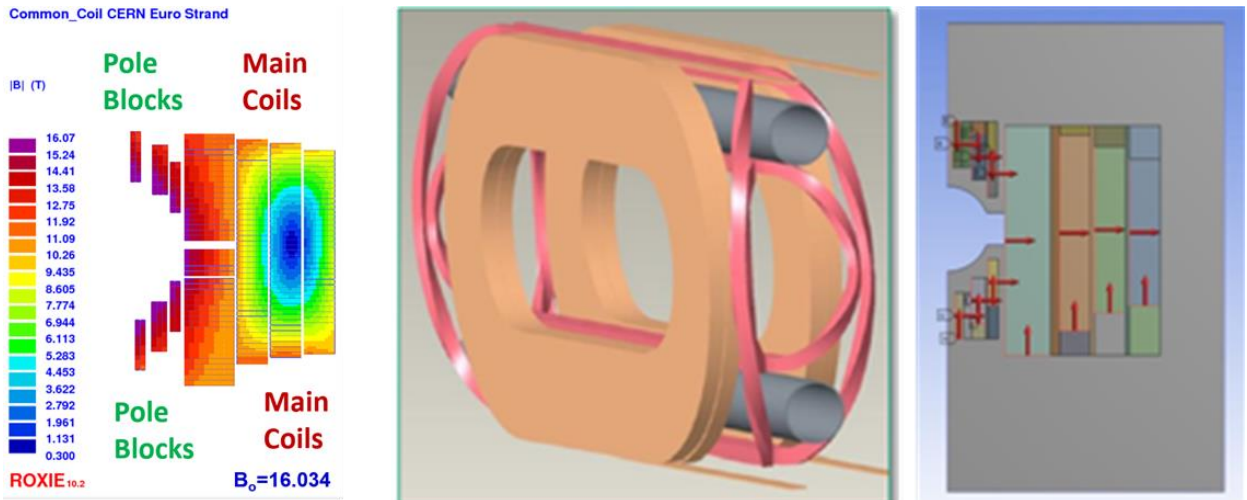

*Fig. 4: Design studies performed for 16 T common coil dipole for FCC by PBL/BNL team. Left: Optimized magnetic design; Center: Layout of the main coils (light brown) and pole coils (pink); and Right: ANSYS model.*

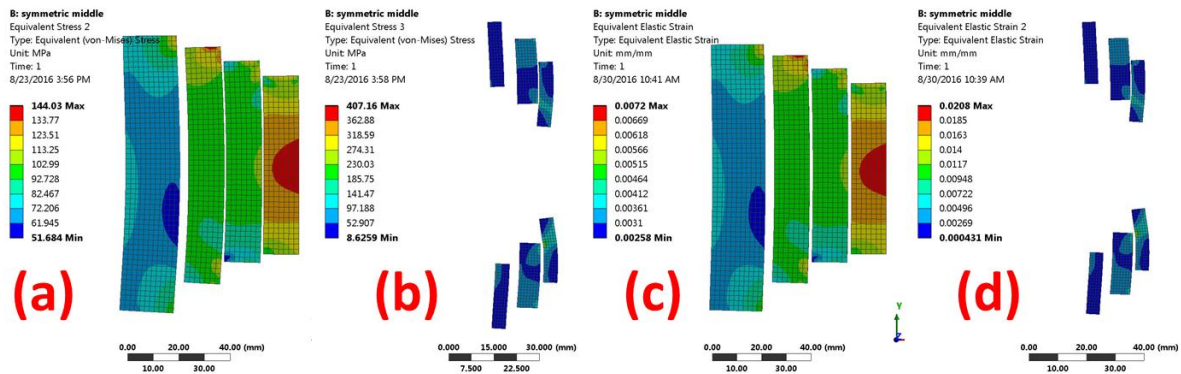

*Fig. 5: Von Mises Stresses in the main coils (a) and in the pole coils (b); Strains in the main coils (c) and in the pole coils (d).*

### **Recent 20 T common coil design study under MDP**

As a part of the US Magnet Development Program, design studies have been started for various design options for 20 T HTS/LTS hybrid collider dipoles. This comparative design study involved a Cosine Theta (CT) dipole, Stress Managed Cosine Theta (SMCT) dipole, Block (BL) dipole, Canted Cosine Theta (CCT) dipole, and the Common Coil (CC) dipole. Preliminary results of this design study were recently presented [17] at the 27th Magnet Technology conference, even though all designs were not optimized to the same level. An interesting finding was that at these field levels the common coil design uses significantly less conductor (particularly much less HTS), as compared to that in the other designs (see Fig. 6 and Fig. 7).

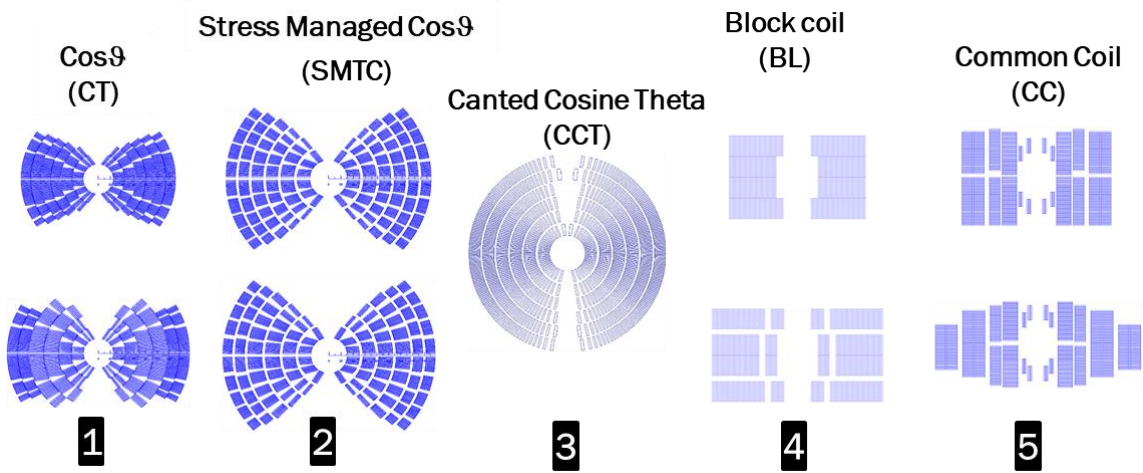

*Fig. 6: Representative early optimization of cross sections for 20 T HTS/LTS hybrid designs with ~15% margin for various designs: (1) Cosine Theta (CT), (2) Stress Managed Cosine Theta (SMCT), (3) Canted and in the pole coils (4), Block design (BL), and the (5) Common Coil (CC).*

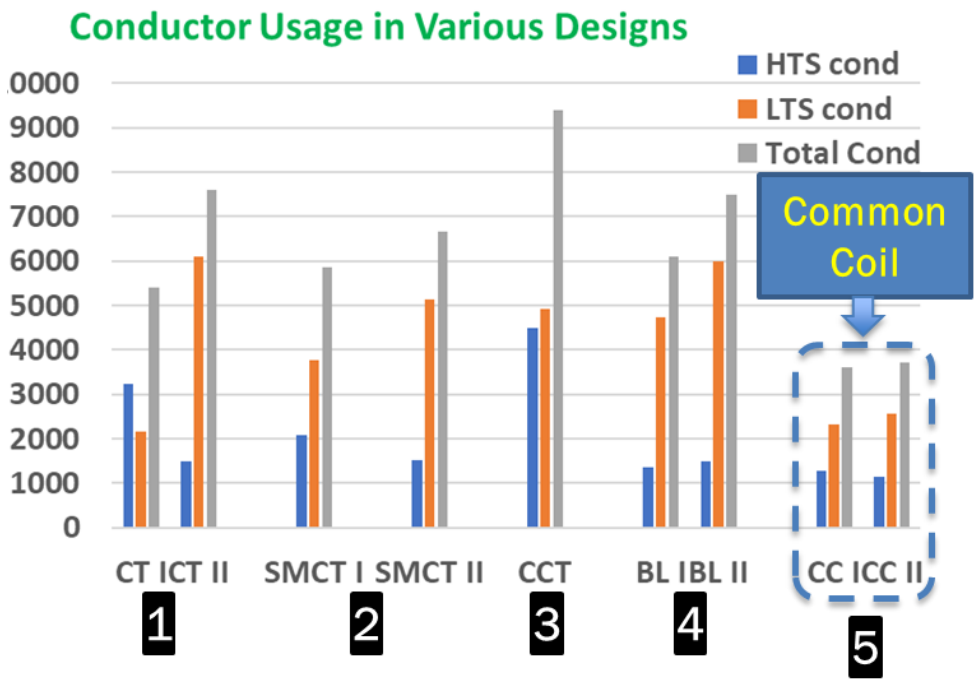

*Fig. 7: Conductor uses in early optimization of cross sections for 20 T HTS/LTS hybrid designs with ~15% margin for various designs: (1) Cosine Theta (CT), (2) Stress Managed Cosine Theta (SMCT), (3) Canted and in the pole coils (4), Block design (BL), and the (5) Common Coil (CC). One can see that the common coil design uses significantly less conductor than the other designs.*

This, as explained at the MDP annual meeting [18], may be due to a significant difference between the low field designs (where the coil thickness is much smaller than the coil aperture) and the high field designs (where the coil thickness is much larger than the coil aperture). A range of common coil design options were examined with a number of layers in the main coils ranging from four (one HTS and three $Nb_3Sn$) to six (one HTS and five $Nb_3Sn$), all using about the same amount of conductor. Further work revealed that all $Nb_3Sn$ layers can be made identical in a reasonably well optimized design (see Fig. 8). This brings a significant savings in the cost of tooling, etc. (such as number of practice coils and spares) which is a large fraction of the cost in an R&D magnet. An initial mechanical design study revealed that it should be possible to design a structure which keeps stresses below 120 MPa and 180 MPa in HTS and $Nb_3Sn$ coils, respectively.

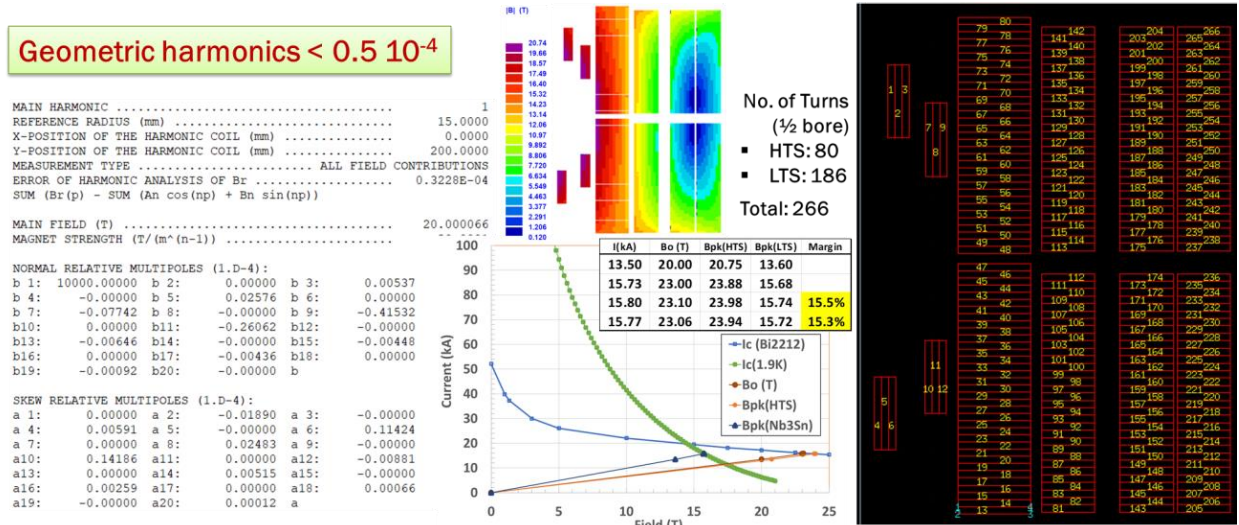

*Fig. 8: A preliminary design of an HTS/LTS hybrid common coil dipole cross-section. Field quality at 20 T is shown in the left, coil cross-section with the field superimposed on the conductor and the operating margins in the middle and the coil cross-section of half of the coil cross-section in upper aperture showing the turn numbers on right.*

**The following is a partial list of the series of tasks**:

- Mechanical analysis of the 20 T HTS/LTS hybrid design
- Provide feedback to the magnetic design for the space needed for the structure between layers and within each layer. Iterate magnetic and mechanical designs
- Develop concepts for assembling the magnet
- Perform 3-d magnetic and mechanical analysis for a 20 T design
- Perform refined mechanical analysis for practical 3-d structures
- Perform quench protection analysis
- Build pole coils and demonstrate them in a proof-of-principle magnet
- Perform cost estimates of R&D dipoles and for large scale series production

## Hybrid common coil R&D dipoles

To reach very high fields, 20 T and above, as discussed in the previous section, HTS/LTS hybrid designs are being examined to save the volume of expensive HTS. R&D on hybrid common coil dipoles is being carried out under various programs. A PBL/BNL hybrid dipole [12] with HTS tape coils in common coil configuration (field perpendicular on the wide face of HTS tape) with the $Nb_3Sn$ Rutherford cable coil, reached a field of 8.7 T in 2016 (see Fig. 9). HTS tape coil for pole coil configuration (field parallel on the wide face of HTS tape) was investigated under US Magnet Development Program [13] reaching a record hybrid field of 12.3 T in 2020 (see Fig. 10). Common coil hybrid dipole with the HTS coil made of CORC® cable running in series with the main coils to reach 13-14 T are also being investigated as a part of the Small Business Technology Transfer (STTR) program with Advanced Conductor Technologies LLC (14) and as a part of the US magnet development program.

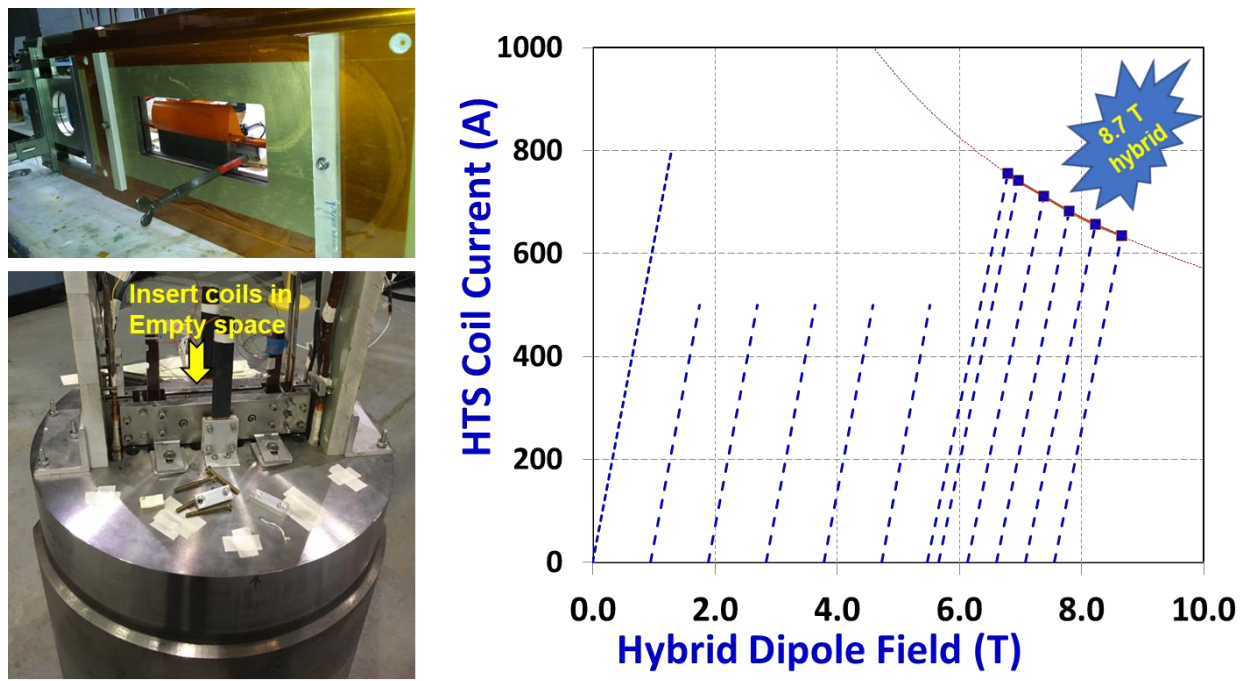

*Fig. 9: HTS/LTS hybrid common coil dipole reaching 8.7 T. Top-left: HTS coil in a frame; Bottom-left: HTS coil integrated with the $Nb_3Sn$ coil; Right: Quench performance of the HTS/LTS hybrid dipole.*

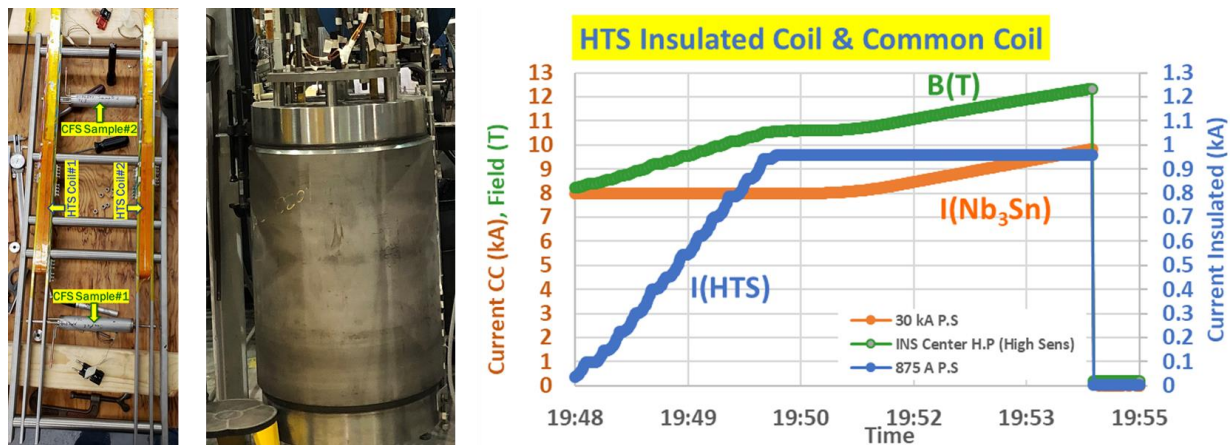

*Fig. 10: HTS/LTS hybrid dipole reaching 12.3 T. Left: Two HTS coils in a frame; Center: HTS coils integrated with the $Nb_3Sn$ coil; Right: Quench performance of the HTS/LTS hybrid dipole.*

## Summary

The common coil geometry provides an alternate design to the conventional cosine theta dipoles. It allows a wider range of conductor and magnet technologies. It also facilitates a low-cost, rapid-turn-around design and R&D program. Some of the benefits of the common coil geometry are listed below:

- Simple 2-d coil geometry for collider dipoles
- Conductor friendly design with large bend radii (determined by the spacing between the two apertures). Less sensitive to conductor degradation.
- 20 T dipole uses significantly less conductor than in other designs
- Efficient segmentation between LTS and HTS coils for HTS/LTS hybrid dipoles
- Mechanically handles well the large Lorentz forces associated with the high fields, creating lower internal strain on the conductor despite large deflections
- Fewer coils (half) because the same coils are shared between two apertures
- Simple magnet geometry and simple tooling leading to lower costs
- Identical design can be used for all $Nb_3Sn$ coils
- Allows both "React & Wind" and "Wind & React" options
- Allows more technology options for insulation, etc.
- Allows rapid-turn-around, low-cost R&D for systematic and innovative studies

## Proposed Program

Based on several unique benefits and opportunities offered by the common coil design as listed in the summary above, it is proposed that the common coil program be a part of a long-term R&D program to develop high field magnets for high energy hadron colliders. Design studies that are being currently carried out under the US Magnet Development Program can be further augmented with the tasks listed on page 7. The general high field magnet technology developed under this R&D will not only be useful to hadron colliders but other very high field magnet programs as well, including the muon collider.